\pdfoutput=1
\documentclass[twocolumn,prb,showpacs,aps,superscriptaddress]{revtex4-1}

\usepackage[english]{babel}
\usepackage[ansinew]{inputenc}
\usepackage{times}
\usepackage{graphicx}
\usepackage{graphics}
\usepackage{amsmath}
\usepackage{amsfonts}
\usepackage{amssymb}
\usepackage{amsbsy}
\usepackage{dsfont}
\usepackage{epstopdf}
\usepackage{makeidx}
\usepackage{subfigure}
\usepackage{color} 
\usepackage{pgf}
\usepackage{bm}
\usepackage{tikz} \usepackage[normalem]{ulem}

\newcommand{\be}{\begin{equation}}
\newcommand{\ee}{\end{equation}}
\newcommand{\bea}{\begin{eqnarray}}
\newcommand{\eea}{\end{eqnarray}}

\makeindex
\begin{document}

\title{Multiscale Modeling of Shock Wave Localization in Porous Energetic Material}

\author{M. A. Wood}
\affiliation{Center for Computing Research, Sandia National Laboratories, Albuquerque, New Mexico 87185, USA}
\author{D. E. Kittell}
\affiliation{Engineering Sciences Center, Sandia National Laboratories, Albuquerque, New Mexico 87185, USA}
\author{C. D. Yarrington}
\affiliation{Engineering Sciences Center, Sandia National Laboratories, Albuquerque, New Mexico 87185, USA}
\author{A. P. Thompson}
\affiliation{Center for Computing Research, Sandia National Laboratories, Albuquerque, New Mexico 87185, USA}

\date{\today}

\begin{abstract}
Shock wave interactions with defects, such as pores, are known to play a key role in the chemical initiation of energetic materials. 
The shock response of hexanitrostilbene is studied through a combination of large scale reactive molecular dynamics and mesoscale hydrodynamic simulations. 
In order to extend our simulation capability at the mesoscale to include weak shock conditions ($<$ 6 GPa), atomistic simulations of pore collapse are used to define a strain rate dependent strength model. 
Comparing these simulation methods allows us to impose physically-reasonable constraints on the mesoscale model parameters. 
In doing so, we have been able to study shock waves interacting with pores as a function of this viscoplastic material response. 
We find that the pore collapse behavior of weak shocks is characteristically different to that of strong shocks.
\end{abstract}

\pacs{}

\maketitle

\section{\label{sec:level1}Introduction}
Many scientific advancements in materials modeling have been enabled by the growing capability of high performance super computers\cite{QMMM09, BulatovBigBig, KadauSci02, PriyaBubble}, but this sort of brute force scaling to discovery falls short for problems that cannot be assigned to a single computational method.
In most cases this is due to an interplay of the underlying physics and chemistry in vastly different length and time domains. 
Computational efforts in shock response of solids places a high demand on the accuracy of the underlying models of mechanical, thermal and chemical response\cite{WoodRCC}.
For example, the relevant length and time scales for shock propagation are proportional to the wave speed which are on the order of km/s (or equivalently nm/ps).
Meanwhile, the plastic deformation and subsequent chemistry occur on much larger length and timescales, which are on the order of $\mu$m to mm and $\mu$s to ms, respectively\cite{MMeyers94}.
This problem is exacerbated by the fact that microstructural features act as nucleation sites for both the plastic and chemical response, requiring that nanometer scale defects be resolved. 

For the present application of interest, it is the thermal nature of chemical initiation in energetic materials which is being studied. 
Energetic materials begin to react at small regions of elevated temperature known as hot spots. 
These regions are formed by the conversion of mechanical into thermal energy at small defects, voids, and other internal features, and are thought to range in size from tens of nm to mm~\cite{bowden1952,field1982,field1992,tarver1996}. 
These microstructure features are present in nearly all high explosives (HEs) in use today, and are be introduced either via sample preparation or by design. 
For example, intentional introduction of hot spot forming defects, i.e. glass microballoons are used routinely to sensitize emulsion~\cite{mendes2014} and liquid~\cite{bouton1999} explosives. 
Natural, inherent, material porosity is known to affect the shock sensitivity of energetic materials~\cite{khasainov1997}, and in limiting cases, the sensitivity of HE powders at low density are seen to be dramatically more reactive~\cite{varesh1996}.
While it is known that the presence of voids in otherwise fully dense HEs will increase their shock sensitivity, there is a lack of consensus within the community as to how shock waves interact with these defects. 
Therefore, an understanding of void collapse is critical for determining initiation thresholds, especially where these materials can be subjected to both intentional and unintentional mechanical activation.

There exist a multitude of reasons why modeling and experiments cannot decouple the physical mechanisms which may lead to hot spot formation: adiabatic compression of trapped gases~\cite{chaudhri1974}, viscoplastic heating~\cite{frey1972}, hydrodynamic jet impingement~\cite{mader1965}, localized shear banding~\cite{austin2015}, and many others~\cite{field1992} all play a role. 
A well-calibrated material model and equation of state might possibly capture all of the different mechanisms; unfortunately, such models are limited by the available thermophysical property data relevant to the high rates of deformation ($>10^4$ s$^{-1}$) and high pressures ($>1$ GPa) associated with pore collapse.
Experimental observations of pore collapse have progressed over the years to include high speed imaging~\cite{bourne1999}, particle image velocimetry~\cite{swantek2010}, and ultrafast spectroscopy techniques~\cite{hambir2001,bassett2017}. 
From these studies, there exist data on the pore collapse time, free surface velocity, and shock viscosity, which are useful for informing the different hot spot mechanisms and for calibrating the material models. 
However, such data does not often appear for the materials of interest, or it may correspond to an unusual sample preparation that does not represent the bulk HE.

Upon collapse an isolated pore will generate a single hot spot, but it is ultimately the collections of interacting hot spots across multiple length and time scales which leads to the build up of a detonation. 
Practical length scales of interest contain hundreds (if not thousands) of pores, which is why scale bridging efforts between multiple modeling and simulation codes are an active area for research\cite{ANLMultiscale, RiceMulti}. 
Two of the more successful approaches of scale bridging of HE initiation appear to be mesoscale simulations~\cite{baer2002,kim2016} and statistically-driven models~\cite{baer2012,nichols2002}, each having their own advantages and limitations.
In general, mesoscale simulations attempt to resolve state variables across a representative volume element (RVE), whereas statistical methods approximate the RVE with an assumed probability distribution function. 
Both approaches are ultimately limited by the accuracy of the physics captured at the mesoscale level. 
The current work is motivated by the need to improve mesoscale simulations with predictive physics and chemical models inferred from some of the largest atomistic simulations to date. 
We show here that materials models for mesoscale modeling can be significantly improved through the merger of experimental data and atomistic simulations.

Specifically, the objectives of the work seek to utilize massively parallel molecular dynamics (MD) simulations in order to deduce various hot spot forming mechanisms in the crystalline explosive hexanitrostilbene (HNS). 
These simulations are based upon a fully reactive\cite{vanDuin2001, Rappe1991} interatomic potential which is implemented in the large scale atomic/molecular code LAMMPS\cite{Plimpton1995}. 
MD simulations of single pore collapse naturally capture all of the different hot spot forming mechanisms (less electronic excitations) and are subsequently used to train a strain rate dependent plasticity model for HNS. 
The constitutive model is then implemented in the continuum hydrocode CTH\cite{mcglaun1990}, and comparisons are made between CTH-LAMMPS at the same pore diameters and shock pressures. 
The results show that molecular dynamics simulations of single pore collpase events can be used to define physically-reasonable mesocale materials models, which can then be used to efficiently study the behavior of samples containing hundreds of interacting pores.
\section{\label{sec:level2}Shock Response of Porous HNS}
\subsection{\label{sec:level2.1}Continuum Approach}
\begin{figure}[t!]
	\includegraphics{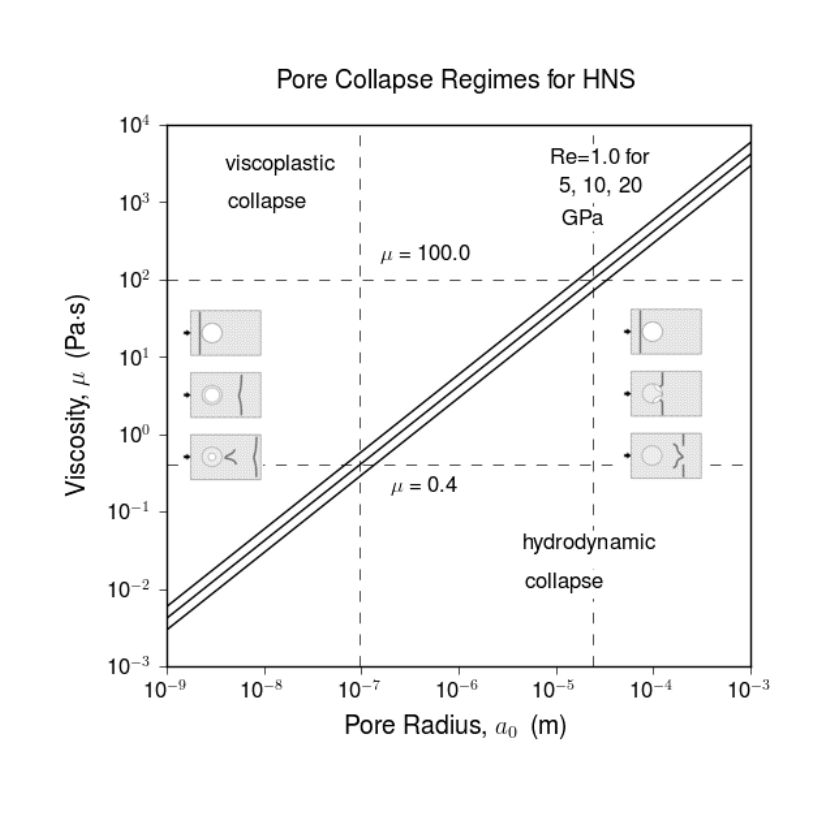}%
	\caption{\label{fig1} Theoretical predictions for the transition from viscoplastic to hydrodynamic pore collapse under shock compression. Estimates for the shock viscosity are from Chou {\it et al.}~\cite{chou1993}}
	\end{figure}

The most common approach for studying shock response of heterogeneous materials is through a continuum mechanics representation of material properties.
Continuum simulations usually follow one of two approaches in the literature: a hydrodynamic model introduced by Mader~\cite{mader1965}, or a viscoplastic model introduced by Carroll and Holt~\cite{carroll1972}, which was later extended by Khasainov {\it et al.}~\cite{khasainov1996}, Butler {\it et al.}~\cite{kang1992}, and Frey~\cite{frey1972}. 
In summary of the different models, the type of pore collapse depends on the ratio of inertial to viscous forces, i.e. Reynold's number. 
For a cylindrical pore geometry, the Reynold's number is given as,
\begin{equation}
Re=\frac{a_0\sqrt{\rho{P}}}{\mu},
\label{Eqn: Reynolds Number}
\end{equation}
where $a_0$ is the initial pore radius, $\rho$ and $P$ are the density and pressure following the shock wave, and $\mu$ is the shock viscosity. 
The shock viscosity is estimated from experimental measurements~\cite{hambir2001} or theoretical calculations~\cite{chou1993}, and is usually orders of magnitude lower than normal values. 
At low Reynold's number ($Re\ll1$), the viscoplastic models assume radial symmetry and incompressible plastic flow behind the shock wave, whereas at high Reynold's number ($Re\gg1$) the compressible flow equations are solved using a hydrocode, and material strength is often neglected. 
Lines of constant $Re=1$ are drawn on a shock viscosity-pore radius plot for HNS, assuming different values for the shock pressure and a range of shock viscosities from Chou {\it et al.}~\cite{chou1993} (see Fig.~\ref{fig1}). 
Insets in Figure \ref{fig1} schematically show these differing pore collapse mechanisms.
Unfortunately, the viscoplastic to hydrodynamic transition is ambiguous across a wide range of pore radii (0.1 to 30 $\mu$m) and both mechanisms may be relevant to HNS initiation. 
Continuum simulations that naturally capture both viscoplastic and hydrodynamic behavior (including material jetting, viscoplastic heating, and shear banding) require a full stress tensor model. 
Only a few such models have been developed to model pore collapse in explosives; one of the most relevant to the current work is the $\beta$-HMX crystal plasticity model of Austin {\it et al.}~\cite{austin2015}.
One key result from that work is rate-independent strength models are incapable of reproducing shear banding and viscoplastic collapse. 
Hence, a simple strength model with plastic strain dependence alone cannot give the desired material behavior.

In this work we employ the strain-rate dependent, Steinberg-Guinan-Lund (SGL) viscoplastic strength model~\cite{steinberg1989} for HNS.
For simplicity, a reduced form of the full SGL model was chosen for tuning the yield strength to the strain rate. 
In summary of this constitutive model for HNS, the stress tensor is decomposed into spherical and deviatoric terms,
\begin{equation}
\sigma_{ij}=-\bar{P}\delta_{ij}+S_{ij},
\label{Eqn: Stress Tensor}
\end{equation}
where the mean pressure, $\bar{P}$, is given by the thermodynamically complete Mie-Gr\"{u}neisen equation of state (EOS) from Kittell and Yarrington~\cite{kittell2016}, and the von Mises yield criteria is assumed to limit the magnitude of the deviatoric stresses,
\begin{equation}
S\le\sqrt{\frac{2}{3}}Y,
\label{Eqn: von Mises Yield Criteria}
\end{equation}
where $S=\vert\boldsymbol{S}\vert=\sqrt{S_{ij}S_{ij}}$ (summation implied over repeated indices throughout) and $Y$ is the yield strength given by,
\begin{equation}
Y=Y_T\left(\dot{\varepsilon}_p,T\right)+Y_A,
\label{Eqn: SGL Model}
\end{equation}
where $Y_T$ and $Y_A$ are the thermal and athermal components, respectively. The value for $Y_A$ is assumed constant, while $Y_T$ contains the strain rate-dependence and is given by the implicit relation,
\begin{equation}
\dot{\varepsilon}_p=\left(\frac{1}{C_1}exp\left[\frac{2U_K}{T}\left(1-\frac{Y_T}{Y_P}\right)^2\right]+\frac{C_2}{Y_T}\right)^{-1},
\label{Eqn: SGL Model YT}
\end{equation}
where $C_1$, $C_2$, $U_K$, and $Y_P$ are material constants and $\dot{\varepsilon}_p$ is the plastic strain rate calculated from the rate of deformation tensor via $\dot{\varepsilon}_p=\sqrt{\tfrac{2}{3}{\dot{e}^p}_{ij}{\dot{e}^p}_{ij}}$. 
In addition, the total rate of deformation tensor is decomposed into an elastic ($e$) and plastic ($p$) component given by Hooke's law and the associated flow rule,
\begin{equation}
\boldsymbol{\dot{e}}^{e}=\frac{1}{2G_0}\overset{o}{\boldsymbol{S}},
\label{Eqn: Hookes Law}
\end{equation}
and
\begin{equation}
\boldsymbol{\dot{e}}^{p}=\lambda\boldsymbol{S},
\label{Eqn: Associated Flow Rule}
\end{equation}
where $G_0$ is the shear modulus calculated from the slope of the Hugoniot elastic limit, $\overset{o}{\boldsymbol{S}}$ is the Jaumann corotational derivative, and $\lambda=\vert{\boldsymbol{\dot{e}}^{p}}\vert/\vert{\boldsymbol{S}}\vert$ is a scalar used to normalize Eq.~\ref{Eqn: Associated Flow Rule}. 
Finally, the SGL model assumes a melt curve of the form,
\begin{equation}
T_m=T_{m0}exp\left[2\gamma_0\left(1-\rho_0/\rho\right)\right]\left(\rho/\rho_0\right)^{-2/3},
\label{Eqn: Melt Curve}
\end{equation}
where $\gamma_0$ is the Gr\"{u}neisen parameter. When temperatures are found in excess of $T_m$, the yield strength is set to zero ($Y=0$).

The SGL model in Eqs.~\ref{Eqn: SGL Model} through~\ref{Eqn: Melt Curve} was implemented in CTH~\cite{mcglaun1990}, a shock physics hydrocode developed by Sandia National Laboratories. 
CTH is used to model multidimensional, multimaterial, large deformation shock wave physics, and employs a fixed Eulerian mesh with Lagrangian and remap solution steps. 
\subsection{\label{sec:level2}Reactive Molecular Dynamics}
In a similar fashion to the distinction between hydrodynamic model forms, molecular dynamics (MD) simulations require the selection of a material model in the form of an interatomic potential (IAP) which is a short-ranged description of atomic energies and forces.
This is the leading approximation in these simulations. 
A wide range of different types of IAP have been developed over the last few decades\cite{DawEAM84, BaskesMEAM92, RappeUFF, Tersoff88, COMB07}.
As a general trend the MD community has been moving toward more accurate and more computationally intensive potentials\cite{PlimptonThompson12, Plimpton1995, Rappe1991}.

A subset of these IAP are known as bond order potentials\cite{REBO02, BrennerIAP} which dynamically calculate per-atom energies and forces as a function of the bonding environment around each atom, allowing for reactions to progress naturally rather than an \emph{ad hoc} approach to chemistry\cite{BrennerAB, HerringAB} used with classical potentials.
The most commonly used of these reactive MD potentials is ReaxFF, which has parameterized force fields  for proteins, ceramics, metal-oxides and organic solids\cite{Senftle2016, vanDuin2001}.
The ReaxFF potential specific for energetic materials has gone through several adaptations since its original implementation by Strachan \emph{et al} \cite{Strachan2003}.
Notably, the original nitramines potential was reparameterized by merging training data with the combustion branch of ReaxFF \cite{Chenoweth2008} as well as including full reaction paths of common energetics (HMX, RDX, PETN) to form the potential reported by Wood \emph{et al} \cite{Wood2014}.
Liu \emph{et. al.}\cite{Liu2011} added a long range, low gradient attractive term to the ReaxFF total energy functional in order to correct for the underestimation of the ambient density.
These density corrected reactive potentials are denoted as ReaxFF-lg for the additional low gradient term (Eq. \ref{lgterm} below).
This model form is employed for the present study on hexanitrostilbene (HNS) and the force field file needed to run these simulations is included as Supplemental Material.
	\begin{figure}[t!]
	\includegraphics{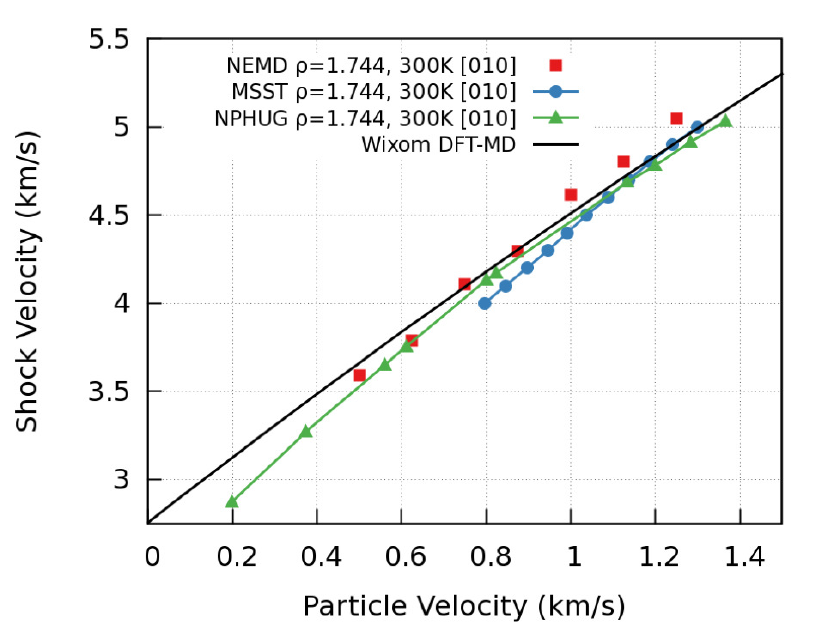}%
	\caption{\label{fig2}Predicted Hugoniot from ReaxFF MD using two equilibrium techniques (MSST, NPHUG) and one non-equilibrium(NEMD) technique in which shock velocity is directly measured.}
	\end{figure}
	
Using the force field from Shan \emph{et. al.}\cite{RayHNSReax}, which is a derivative of the combustion and energetics merger, corrections to the low gradient term were needed in order to accurately predict the ambient density of HNS.
This was done by adjusting the constant term, $d$, term in equation \ref{lgterm} from its initial value of 1.0 to 0.9125 while holding the other constant term, $C_{ij}$, at 0.55.
This value was optimized by running MD simulations coupled to a thermal reservoir at 300~K and a pressure reservoir at 1~atm and comparing the predicted density to the target experimental density of $1.744 \frac{g}{cm^{3}}$.
\begin{equation}
\label{lgterm}
E_{lg}=\sum\limits_{i\neq j}^N\frac{C_{ij}}{r^{6}_{ij}+dR^{6}_{e,ij}}
\end{equation}

To confirm that these changes to the reactive potential do not significantly alter the behavior of HNS, we compute the $U_{s} - U_{p}$ shock Hugoniot, a property that is central to results presented here, using a variety of MD methods.
In Figure \ref{fig2}, the DFT-MD data from Wixom \emph{et. al.}\cite{WixomMattsonSAND} is plotted as the solid black curve along with two equilibrium methods, Multi-Scale Shock Technique (MSST)\cite{ReedPRL03}, and Constant Stress Hugoniotstat (NPHUG) \cite{RavelPRB2004}.
Also included in Figure \ref{fig2} are direct measurements (NEMD) of the shock velocity from a [010] directed single crystal shock experiment using the simulation details outlined in Shan \emph{et. al.}\cite{ShanPRB16}.  
Each of these methods results in a good agreement with the DFT reference data in the range of piston velocities ($0.5-1.25~km/s$) of interest here. 
Details about these two methods are contained in the Supplementary Material.

With a properly trained reactive force field in place, we now turn our attention toward the dynamics of shock induced pore collapse.
It is worth noting that the extra cost associated with running the MD simulations with the reactive versus a non-reactive IAP is necessary in order to capture shock induced chemistry and the natural evolution of the hot-spot in the energetic material\cite{Wood2015}.
\section{\label{sec:level3}Scale Bridging Methods}
\subsection{\label{sec:level3.1}SGL Strength Model Training}
\begin{figure}[t!]
	\includegraphics{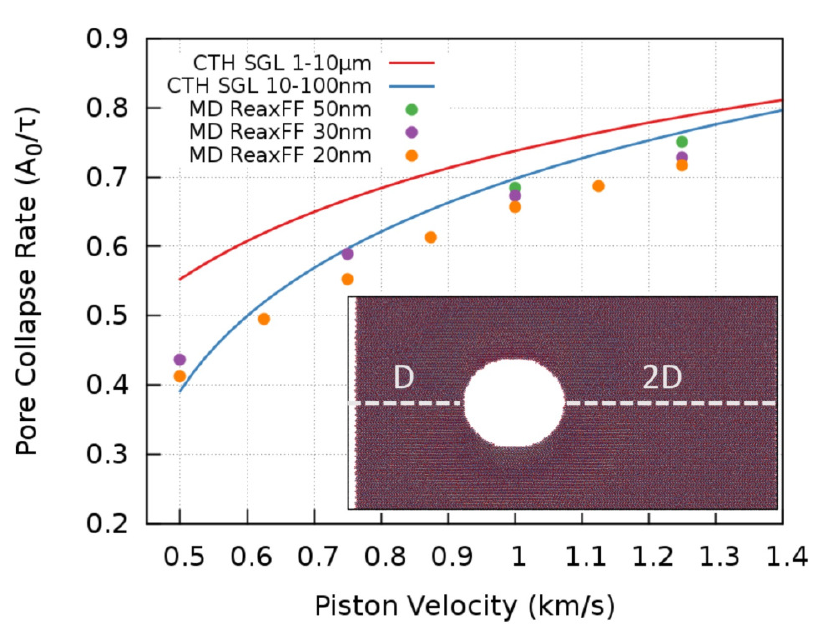}%
	\caption{\label{fig3} Shared metric for the MD and CTH viscoplastic shock response. Also pictured is the cell geometry that was used to explore pore collapse mechanisms as a function of pore diameter (D) and piston velocity. The same cell geometry is used in continuum simulations discusses later.}
	\end{figure}

In order to properly calibrate the SGL strength model, both codes must share an observable quantity that represents the shock response of the porous material. 
This quantity should be sensitive enough to capture the characteristics of a hydrodynamic versus viscoplastic shock response within the limited time and length scales accessible to MD.
Furthermore, the training metric should directly or indirectly exercise many, if not all, of the free parameters in Eqs. 4 through 8 in order to ensure a uniqueness in the fitted solution.
To satisfy these constraints, we have designed the scale independent simulation geometry shown as an inset to Figure~\ref{fig3}, and will use the pore collapse rate as the shared metric between either simulation code. 
In this simulation setup, the material is impacted at the left surface with a fully supported piston of variable velocity.
For the MD simulation of this geometry, the shock is always directed along the [010] crystallographic direction in the (100) plane.
More details about the initial conditions for either code can be found in the Supplemental Material.
To simplify the analysis for every pore diameter and piston velocity pairing, we define a characteristic time ($\tau=D/U_s$), where $D$ is the original pore diameter and $U_s$ is the shock velocity for the given piston velocity, which represents the time for the shock to travel a single pore diameter.  

The slopes of normalized pore area ($\hat{A}=\frac{A(t)}{A(t=0)}$) versus $t / \tau$ are collected from the NEMD runs as the set of points in Figure \ref{fig3}. 
While the pore size does have a noticeable effect on the collapse rate, the main driver behind the transition from purely viscoplastic $(\lim_{U_p \to 0} \frac{\partial \hat{A}}{\partial \tau})$ to hydrodynamic $(\lim_{U_p \to \infty} \frac{\partial \hat{A}}{\partial \tau})$ is the piston velocity that drives the shock wave. 
This can be seen by comparing the range of pore collapse rates at a given piston velocity to the range of values for a given pore size. 

To generate these data from the continuum code, a mesh resolution was achieved using 400 zones across the pore diameter, and the domain was held constant at 1600 by 1200 cells; symmetry conditions were imposed on the impact wall and periodic boundary conditions in the lateral direction to mimic the simulation geometry used in the atomistic code.

The fitted SGL model parameters are summarized in Table~\ref{tab1}, and were found by matching the training metric shown in Figure 3. 
The use of a Latin hypercube sampling (LHS) algorithm~\cite{mckay1979} and $>$5,000 simulation runs were required to fine-tune the model fit; however, further improvement could be made to strength correct the EOS.
Despite this, the fitted parameter values are physically realistic; for example, the value of the drag coefficient $C_2$ is similar to the value obtained from experimental measurements of void collapse in polymethyl\-methacrylate (PMMA)~\cite{hambir2001}, and the Peierls stress for HNS is greater than the yield stress of 140~MPa, but less than the shear stress of 5686~MPa.

\begin{table}[b]
	\centering
	\caption{\label{tab1}HNS parameter values for the reduced SGL model.}
	\begin{tabular}{lll}
	\hline
	\hline
	\noalign{\vskip 1mm}
	Parameter & Value & Fitted\\
	\noalign{\vskip 1mm}
	\hline
	\noalign{\vskip 1mm}
	Initial Yield Strength, $Y_A$\hspace{10mm} & 140 MPa\hspace{10mm} & no\\
	Shear Modulus, $G_0$ & 5686 MPa & no\\
	Melt Temperature, $T_{m0}$ & 588 K & no\\
	Gr\"{u}neisen Parameter, $\gamma_0$ & 1.625 & no\\
	\noalign{\vskip 1mm}
	\hline
	\noalign{\vskip 1mm}
	Pre-Exponential factor, $C_1$ & 2.025e11 s$^{-1}$ & yes\\
	Drag Coefficient, $C_2$ & 1.125 Pa-s & yes\\
	Peierls Stress, $Y_P$ & 1114 MPa & yes\\
	Activation Energy, $U_K$ & 1576 K & yes\\
	\noalign{\vskip 1mm}
	\hline
	\hline
	\end{tabular}
\end{table}
One of the main advantages of training the strain rate dependent (SRD) SGL model to reproduce the MD shock response of HNS is the larger range of defect sizes and shock strengths that can be accurately simulated with the continuum shock physics approach.
To demonstrate this, Figure 4 collects several thousand individual CTH simulations that span a wide range of shock pressures and pore sizes of the same simulation cell geometry used during training.
The color axis in this figure is the scaled pore collapse rate that was defined in the discussion of Figure 3. 
Additionally, the measured pore size distribution of HNS\cite{DammHNS} is shown on a common horizontal axis having a mean and standard deviation of $0.9 \pm 7 \mu m$ that matches experimentally observed microstructures. 
As shown in Figure 4, the primary factor controlling the transition from viscoplastic (blue, region A) to hydrodynamic collapse (red, region B) is the input shock pressure,  i.e. moving from region A to B. 
However, there does exist a size effect that is most apparent at lower shock pressures, i.e., moving from region A to region C. 
	\begin{figure}[t!]
	\includegraphics{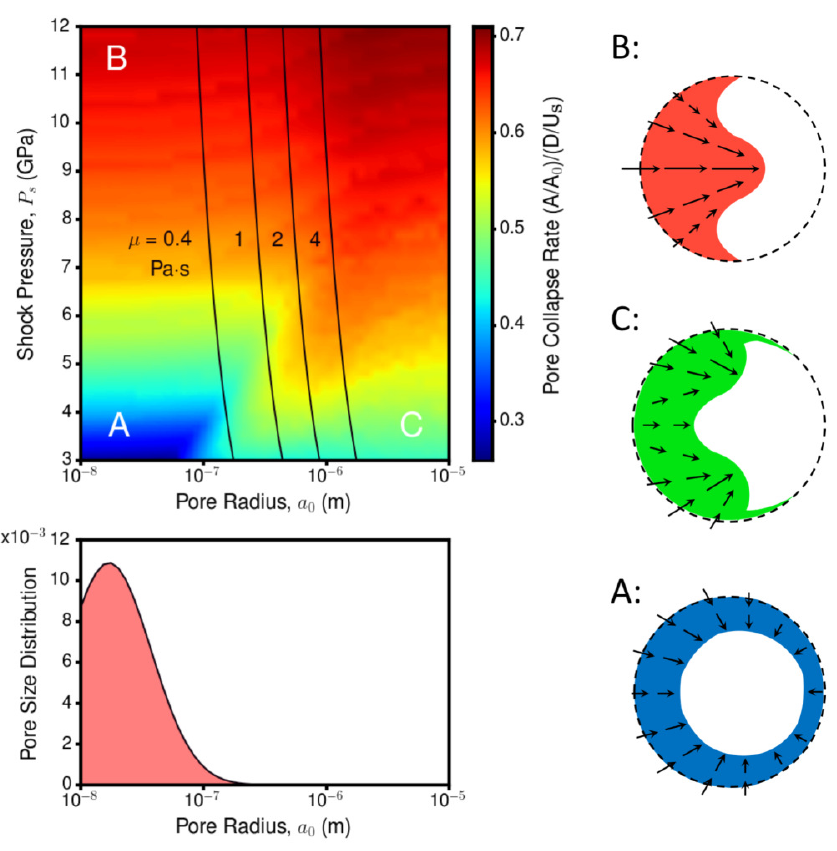}%
	\caption{\label{fig4} {\bf Top Left} Color map of characteristic pore collapse rates calculated using the calibrated SGL mesoscale model for physically relevant ranges of pore size and shock strength. Low (blue to green) and high (yellow to red) collapse rates correspond to conditions of viscoplastic and hydrodynamic shock response, respectively. {\bf Bottom Left} Experimentally measured pore size distribution in HNS\cite{DammHNS}. The majority of the pores lie in the range $10^{-8}-10^{-7}$ m, where pore collapse behavior is sensitive to shock pressure. {\bf Right} Schematics of the pore surface morphology during collapse.  Viscoplastic pore collapse (i.e. region A) occurs for small pores and low shock pressure. Hydrodynamic collapse occurs at high shock pressures and/or large pores (i.e. region B)}
	\end{figure}	

It is also possible to use the size transition between region A and C in Figure 4 as a criterion to estimate the shock viscosity. 
Solid black lines of constant $Re=1$ for Newtonian fluids with viscosity, $\mu$, are plotted in shock pressure-pore radius space from a manipulation of Eq.~\ref{Eqn: Reynolds Number}. 
In contrast to the wide range of viscosities shown in Figure \ref{fig1}, we predict a much smaller range of shock viscosities to define the viscoplastic-hydrodynamic transition, on the order of $1 Pa\cdot s$. 
In turn, these new estimates of the shock viscosity, in conjunction with the data in Figure \ref{fig1}, provide a \emph{critical pore size} the separates viscoplastic or hydrodynamic style of pore collapse. 
From the calibrated SGL model, we find that pore sizes in the range 0.1--0.5 $\mu$m define this transition region.
In the next section, we will explore these limiting cases of shock response with both codes by comparing qualitative and quantitative measures that were not used as training points for the SGL strength model.
\subsection{\label{sec:level3.2}Void Collapse}
To properly test the accuracy of the strain rate dependent SGL model, another set of metrics common to both codes is needed that are not directly used as training.  
In the previous sections we focused on defining the characteristics of viscoplasic and hydrodynamic styles of pore collapse simply through the rate of collapse.
In this section we detail these differing mechanisms by analyzing the temperature and strain fields around the collapsed pores.
The aim here is to gauge the ability of CTH to match the large-scale MD prediction of the same simulation geometry. 
To exemplify the improvements to CTH, we show the MD results for experimentally relevant pore sizes (0.1 $\mu$m) alongside the strain rate independent~(Hydro) and dependent~(SGL) forms of the HNS strength model. 
Of course, we cannot expect that the agreement between MD and CTH be exact, but rather we aim to capture the main features of the strain and temperature fields.
	\begin{figure*}[t!]
	\includegraphics{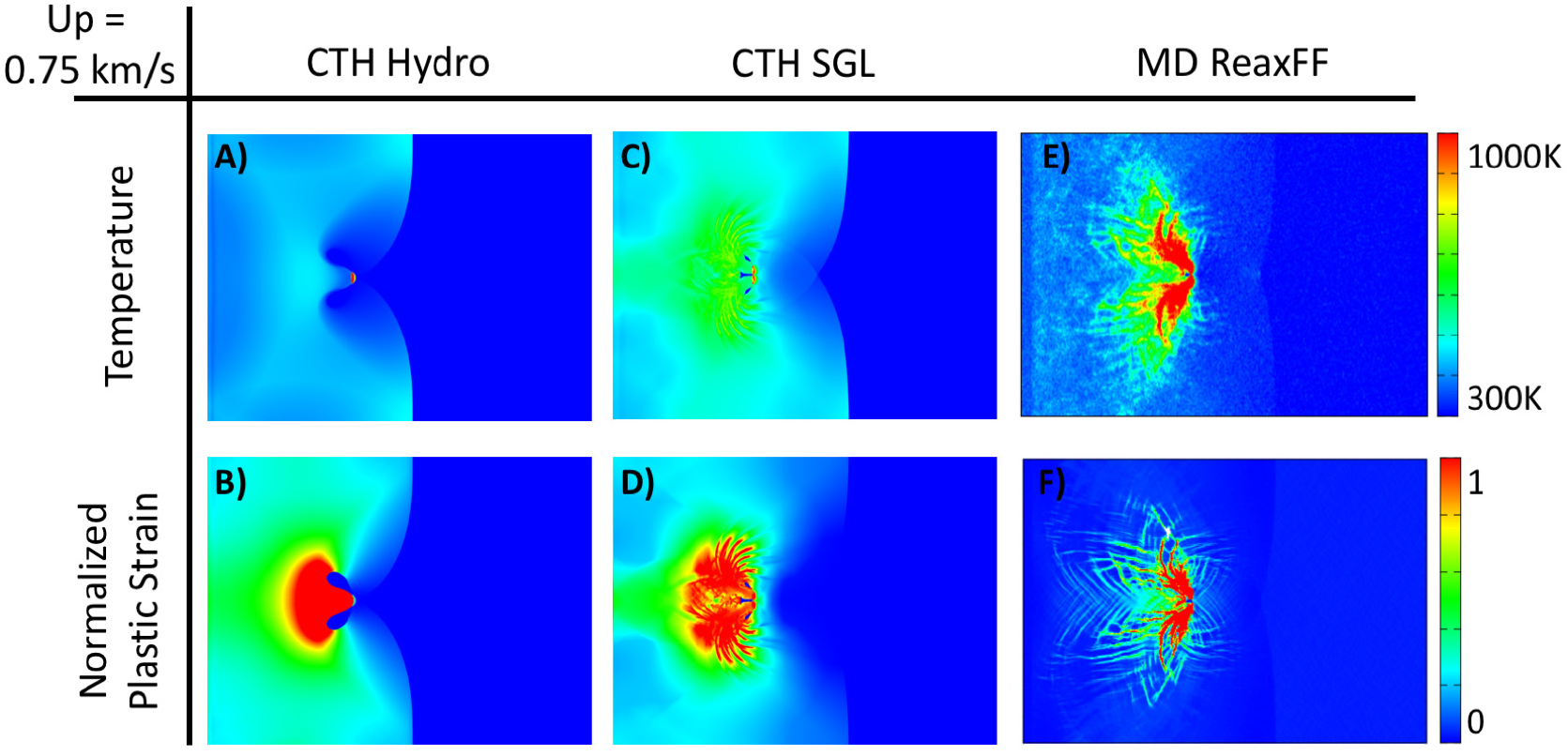}
	\caption{\label{fig5} Comparison of the $U_{p}=0.75$~km/s shock response of porous HNS for a single 0.1~$\mu$m pore. Temperature predictions from the CTH hydro model in regions of high plastic strain are noticeably colder than the strain rate dependent SGL model, which better captures the correlation of strain and temperature observed in the MD simulations.}
	\end{figure*}
	
Where appropriate, the use of CTH to model the shock response of HNS is strongly advantageous because the time to solution in CTH is many orders of magnitude faster than MD.
For example, using the geometry shown in Figure \ref{fig3}, a MD simulation cell with a 0.1~$\mu$m pore contains 20.6 million atoms and, for the slowest piston velocities, requires 200ps elapsed time before the shock reaches the free surface.
A single one of these detailed MD simulations requires a significant allocation of computing resources often unavailable or unfeasible given the amount of compute time needed. 
Utilizing the Intel Knights Landing hardware partition on the Trinity machine at Los Alamos National Lab, these MD simulations required approximately $25\cdot10^{6}$ cpu$\cdot$hrs to complete. 
Further computational details and timing data for these large ReaxFF MD runs can be found in the Supplemental Material.
In contrast, the same simulation can be run with the calibrated SGL strength model through CTH in only 20 cpu$\cdot$hrs, a reduction by a factor of $1.25\cdot10^{6}$ in time to solution.

Figure \ref{fig5} collects the local temperature and plastic strain for a shock that was generated with a piston moving at 0.75~km/s.
Each panel in this figure shows the simulation frame just before ejecta impact.
For both quantities, the results are shown for both CTH strength models and the large-scale MD simulation.
Inspecting the shape of the ejecta that is formed, it is clear that the strain rate independent strength model, Fig. \ref{fig5} A and B, predicts a strong fluid-like jet that originates from the centerline of the pore.
In contrast, the strain rate dependent CTH simulation, Fig. \ref{fig5} C and D, and the MD prediction, Fig. \ref{fig5} E and F, show that the ejecta is formed equally at two axially offset locations of the pore surface that flow toward the centerline of the pore.
This phenomenon is best exemplified by the regions of highest plastic strain which in turn have the highest local temperatures in the MD simulation.
A detailed look at the material flow around the pore surface is given in the Supplemental Figure \ref{Sfig2}. 
This is an important distinction between the strength models in CTH because the purely hydrodynamic pore collapse is seen for all pore sizes and shock strengths for the strain-rate independent form.
Meanwhile the SGL model is capable of capturing both the strong jet formation in the strong shock limit and smoothly transition toward a radially symmetric pore collapse in the weak shock limit, giving it a much larger range of applicability of the space shown in Figure \ref{fig1}.
\subsection{\label{sec:level3.2}Hot-Spot Formation}
The second outside metric that we will compare between the atomistic and continuum methods is aimed at characterizing hot-spots that result from viscoplastic pore collapse. 
Specifically, we aim to provide a physical understanding of how much heat is generated from the ejecta impact versus the shear banding and other plastic material flow around the pore.
In the limiting case of a purely viscoplastic pore closure, the ejecta formation will be suppressed due to irreversible plastic deformation around the pore. We have confirmed this behavior for the weakest piston velocity impacts with the MD data we have generated here.

\begin{figure}[t!]
	\includegraphics{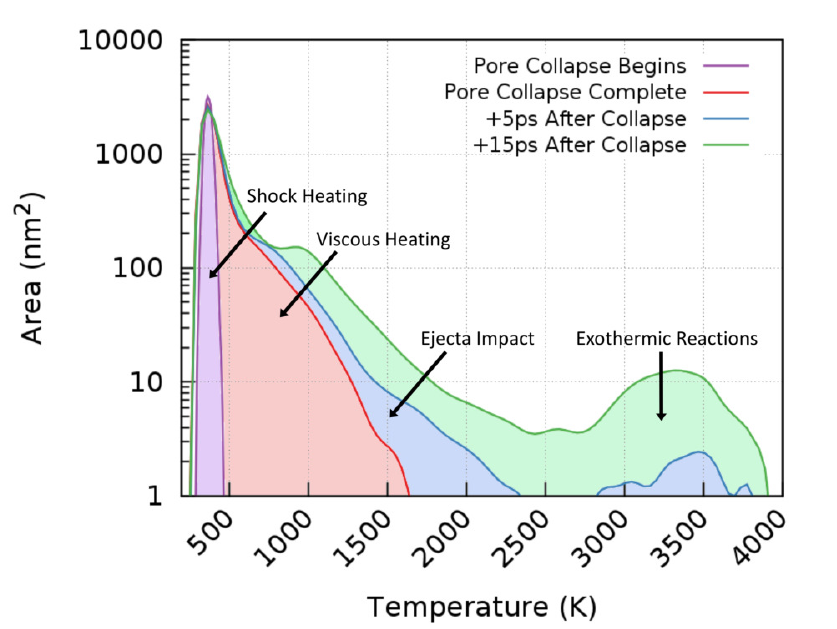}
	\caption{\label{fig6} Temperature histograms at several critical stages in the $U_{p}=0.75$~km/s shock response of HNS with a single 0.1~$\mu$m void via large-scale MD. Labelled arrows indicate the dominant heating mechanism for the corresponding colored region of the temperature distribution.}
	\end{figure}

Figure \ref{fig6} shows the distribution of temperatures that are present in the MD simulation of a $U_{p}=0.75$~km/s shock at various stages during its compression.
Each data series has local temperatures averaged in $1 nm^{2}$ square pixels in the viewing plane shown in Figure \ref{fig5} which is then collected as a histogram with bin width of 8.5~K. Only the material that has been compressed by the leading shock wave is included here.
Therefore, there is no peak in these histograms corresponding to $T=300K$ of the un-shocked material.
The initial heating of the sample is caused by the shock front compressing and adiabatically heating the material. This temperature distribution is shown in purple and labeled as \emph{'Pore Collapse Begins'} in Figure \ref{fig6}.   
For reference, the red region in \ref{fig6} labeled \emph{'Pore Collapse Complete'} corresponds to panel E) in Figure \ref{fig5}.
At this time during the simulation, there is extra heat generated from the viscous flow of HNS which comprises a very large area of the sample with temperatures ranging from 500-1500K.
Each data series after the pore has fully collapsed incorporates multiple heating mechanisms, namely the collisions of ejected molecules and energy release due to chemical reactions. 
To show the importance of these additional heat sources, the temperature distribution for 5ps and 15ps after the pore collapse has completed is shown in blue and green, respectively, in Figure \ref{fig6}.
Relative to the viscous heating regions, these additional sources represent a small mass fraction of simulated material, but are placed at much higher temperatures 1500K-4000K.
However, this small volume of rapidly reacting material contributes strongly to the growth of the hot-spot as exemplified by the bolstered peak near 3500K after 15ps post pore collapse. 
For comparison, we have included the same data for a much more viscoplastic case ($U_{p}=0.50km/s$) in the supplemental material where it can be seen that the ejecta impact shoulder on the viscous heating is much smaller and the exothermic reactions occur at a much slower pace than Figure \ref{fig6}.

Less the adiabatic heating from the shock front, the shaded area indicating viscous heating has a mean temperature of 860K, and given the integrand of this segment, would equate to an area of 226 $nm^{2}$ at this  temperature.
The ejecta impact contributes significantly less to the overall hot-spot that is formed, its mean temperature being 1240K, which can be equated to an area of just 27 $nm^{2}$ at this mean temperature.
Comparing these two mechanisms, the ejecta impact at this (moderately low) piston velocity contributes a factor of six less $Area \cdot Temperature$ than that of the viscous heating mechanism.
In addition, the temperature histograms in Supplemental Figure \ref{Sfig2} show that this ratio of \emph{hot-spot potency} favors viscous heating by a factor of twenty-five. 
This trend continues up to piston velocities of 1.25km/s where the ratio decreases down to 1.78, but still in favor of viscoplastic heating being the dominant mechanism. 
Of course the time delay used to capture the ejecta impact is somewhat arbitrary and needs to be carefully chosen since HNS begins to react promptly after the impact occurs\cite{sidecomment1}.  

\begin{figure}[t!]
	\includegraphics{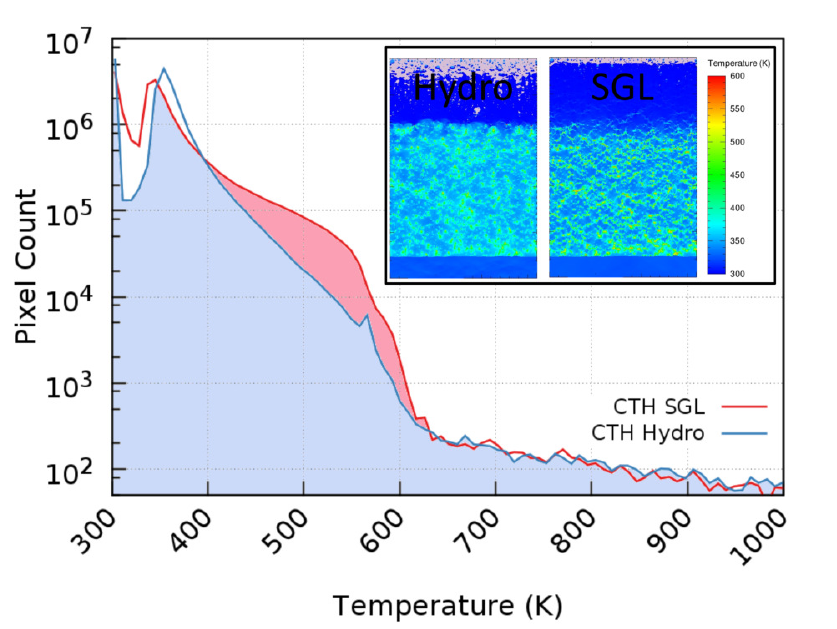}
	\caption{\label{fig7} Temperature histograms for both CTH strength models: CTH Hydro (red); CTH SGL (blue). Insets are snapshots of the simulation cell in both cases. The CTH SGL model informed by MD results shows a significant increase in heat generated from shear deformation.}
	\end{figure}

Up to this point, the discussion of hot spot mechanisms has been focused on idealized pore geometries, but with the SGL strength model in CTH we are able to address much more realistic shock responses of HNS by providing mesoscopic porous microstructures as input geometries. 
The insets to Figure \ref{fig7} show a $15\mu m$ by $25 \mu m$ slab of HNS with a pore size distribution that is artificially generated such that it matches experimentally observed microstructures.
The sample is shocked at the bottom surface with a piston moving at 0.60km/s for both the Hydro and SGL strength models, each simulation snapshot and temperature histogram corresponds to when the shock wave reaches the free (top) surface.
While Figure \ref{fig7} is the culmination of heat generated from many collapsed pores, the SGL model clearly shows additional heat generated in the 400-600K range that is due to viscoplastic heating, see also Figure \ref{fig5} A),C).
Accurately predicting hot spot temperatures is a crucial step in predicting the detonation performance of HNS and many other energetic materials, and the SGL model presented here is a significant advancement for these simulation methods. 
\section{\label{sec:level6}Conclusions}
The work presented here demonstrates a key advancement in multi-scale modeling of HE initiation in that atomistic simulation results are directly used to refine material strength models used in continuum codes. 
These continuum codes have already built upon experimental results for their equations of state and heat capacity models but there exist several gaps in knowledge about hot spot forming defects, which is where this effort is focused. 
We determined that a strain-rate dependent strength model for HNS was needed in order to simultaneously capture the heat generated from plasticity as well as ejecta impact during pore collapse. 
A limited set (so as to avoid over fitting) of training data was constructed through medium (approx. $10^5$ atoms) and extreme scale ($> 10^7$ atoms) reactive MD simulations that capture the collapse of isolated voids.
These training data spanned a wide range of shock strengths and pore sizes which was necessary to exercise the limits of material behavior from purely viscoplastic to purely hydrodynamic. 
As the style of pore collapse changes (see Figure 4), we found that the primary source of heat around the collapsing pore also changes.
This transition from ejecta impact heating for strong shocks ($>6$ GPa in HNS) and plasticly driven heating for weak shocks is captured naturally in MD, as well as within CTH with the trained SGL strength model developed here.
To date, there has been a necessary focus on the response of materials under very strong shock conditions because these conditions best informed efforts to predict detonation performance.
In contrast, and more recently, safety of energetic components has been the focus of many research efforts which includes cases far from the design condition, like low velocity impacts that were the focus of this work. 
The advancements made in this contribution enable continuum mechanics simulations of initiation under much weaker mechanical impacts.
In addition, the improved fidelity of this HNS material model will enable predictive simulations of the shock to deflagration, and shock to detonation transition from simulations that include microstructure detail, a capability that is currently lacking in the field of shock physics. 
Extensions of this work will be targeting ensembles of microstructure features and their role on the ignition and growth process that yields a sustained detonation wave. 
\begin{acknowledgments}
We would like to thank Ryan Wixom and Yasuyuki Horie for valuable discussions throughout this work. 
Computing time on the Trinity supercomputer at Los Alamos National Lab was provided by the ATCC-3 campaign.
This document is approved for public release under SAND2017-12295 J.
Sandia National Laboratories is a multi-mission laboratory managed and operated by National Technology and Engineering Solutions of Sandia, LLC., a wholly owned subsidiary of Honeywell International, Inc., for the U.S. Department of Energy's National Nuclear Security Administration under contract DE-NA0003525.
\end{acknowledgments}

\widetext
\begin{center}
\end{center}

\widetext
\newpage
\begin{center}
\textbf{Supplemental Material}
\end{center}
\setcounter{equation}{0}
\setcounter{figure}{0}
\setcounter{table}{0}
\setcounter{page}{1}
\makeatletter
\renewcommand{\theequation}{S\arabic{equation}}
\renewcommand{\thefigure}{S\arabic{figure}}
\renewcommand{\bibnumfmt}[1]{[S#1]}
\renewcommand{\citenumfont}[1]{S#1}

Within the LAMMPS molecular dynamics code there are several options to calculate the Hugoniot of a defect free sample. 
Two of these methods, MSST and NPHUG, are computationally efficient because they only require a small periodic cell which can simply be a unit cell of the material. 
Each of these methods adjust the equations of motion of the atoms in order to meet the conservation of mass and momentum across a shock front. 
In practice, the MSST protocol takes a target shock velocity as input and will calculate the change in volume and temperature needed to meet the Hugoniot jump conditions. 
Conversely, NPHUG will take a target pressure as input to calculate the same quantities but this method can be thought of as a modified Nose-Hoover barostat that will converge on the Hugoniot state point. 
In our MSST simulations shown in Figure 2, a cell mass-like parameter of 0.1, artificial viscosity of 0.0 and an initial temperature reduction of 0.01 were used; these parameter vary for each material and care needs to be taken in their selection. 
Shock velocities between 4.0 and 5.0km/s were selected for MSST, and pressures between 1.0 and 13.0 GPa were used for NPHUG. 
The last method of determining the particle, shock velocity Hugoniot points are through direct simulation of a shock wave passing over a perfect crystal of HNS. 
Here a unit cell of HNS is replicated 1x80x36 by times in the [100], [010] and [001] directions respectively and is impacted on the (010) surface. 
The position of the shock front is located by the change in density.
In this direction there is only a single wave structure which is to say it is a plastically overdriven wave. 
In other crystallographic directions the shock wave was observed to have a split wave with a leading elastic and trailing plastic wave. 
Each of these methods agree well with the reference DFT data, giving us confidence that our changes to the low gradient term in the ReaxFF parameterization have not perturbed the shock response of HNS. 

The Trinity supercomputer at Los Alamos National Lab is a Cray machine that contains two partitions of distinct multiple integrated core hardware environments, one of 9,408 Intel Haswell nodes and another of 9,200 Intel Knights Landing nodes each utilizing the Cray Aries Dragonfly interconnect. 
Haswell compute nodes have two, 16-core sockets per node, and can use up to two hyperthreads per core giving a peak predicted performance of 11.0Pflops when using all 602,112 available cores. 
Knights Landing(KNL) compute nodes have a single socket with 64 cores (2 per tile) which can handle up to 4 hyperthread tasks per core giving a peak predicted performance of 42.2Pflops when using all 2,355,200 available cores.
Detailed information about LAMMPS performance on these arcitechtures can be found from Moore {\emph et. al.} \cite{MooreTrinity}.
During the open science testing period on the KNL partition of Trinity, we were able to build and run KOKKOS compiled versions of LAMMPS designed to take advantage of this unique hardware. 
Typically these production jobs were run with ~250 atoms/processor where we would expect ~2 MD timesteps/second where a 0.1fs timestep increment was used. 
For MD simulations using the ReaxFF force field, we believe these simulations are the largest (20.6 million atoms for  $>$1million timesteps) to ever be reported on.
At these scales we observed that the current implementation of QEq, the charge updating scheme, was as computationally intensive as the calculation of pairwise forces on each of the atoms, exposing a potential bottleneck at even larger scales\cite{NiklassonPRL08}.
 \begin{figure}[]
\includegraphics{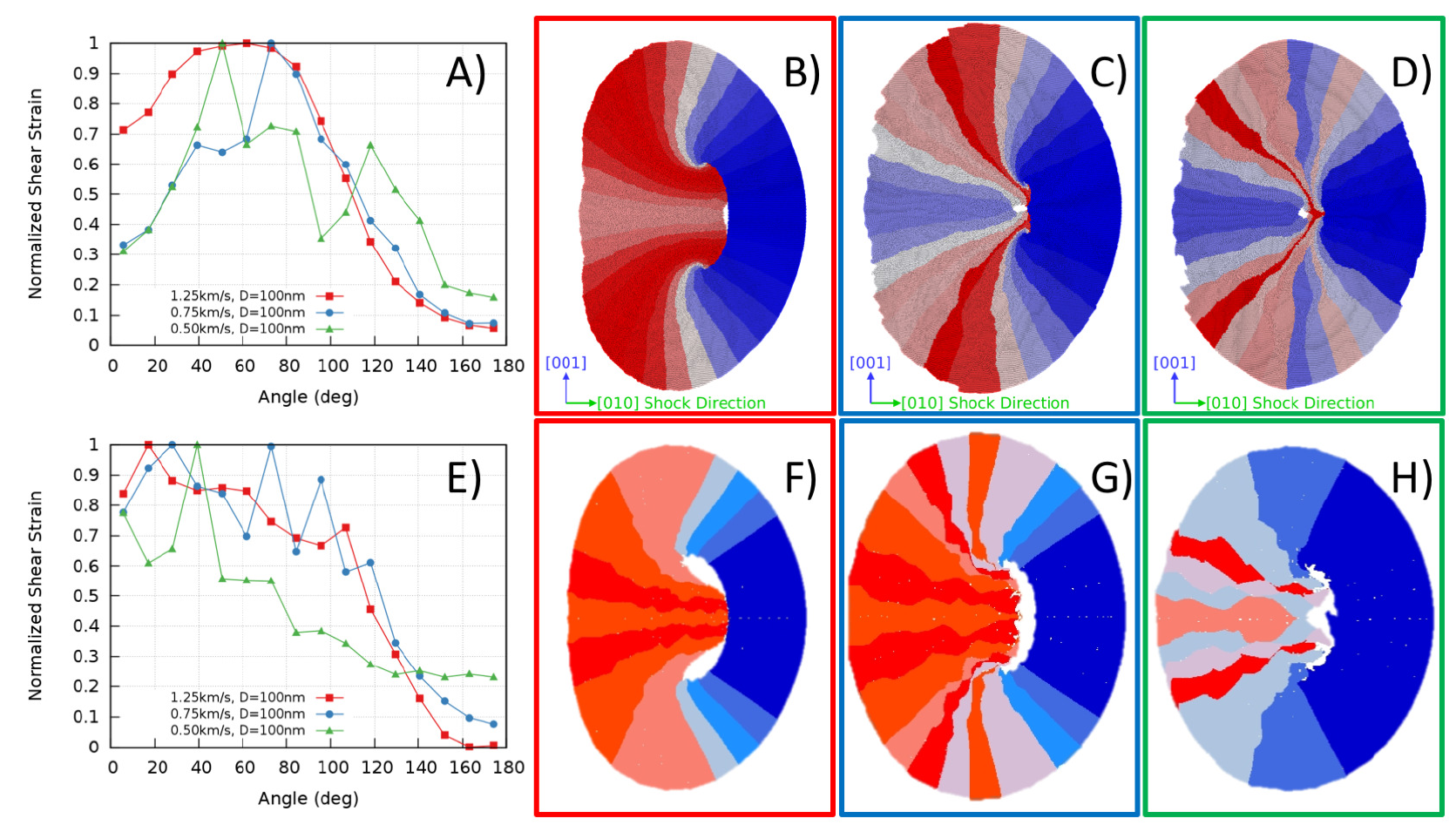}%
\caption{\label{Sfig1} {\bf A)} Angular strain distribution around fully collapsed 100nm pores. As the piston velocity is increased, the peak strains increase but also become more evenly distributed around the pore as seen by the 0.50km/s and 0.75km/s data at angles greater than 110 degrees from the upstream surface. {\bf B)} Strain distribution mapped onto the area around the pore surface for a piston velocity of 1.25km/s; fluid-like ejecta is seen to form predominately from the upstream surface. {\bf C)} Intermediate piston velocity of 0.75km/s. Strain field is more evenly spread out around the pore surface. {\bf D)} Low piston velocity of 0.50km/s showing a purely viscoplastic pore collapse where material from all angles around the pore merges on a central point inside the pore. Panels E) though H) display the same results predicted from CTH using the MD trained SGL model.}
\end{figure}

It is worth noting that the point of ejecta impact on the downstream pore wall is roughly the same between all simulation cases shown in Figure \ref{fig5}.
This is important because there is a strong secondary shock wave generated from the ejecta impact that will subsequently heat the material downstream of the pore when the leading and secondary shock waves overlap\cite{ShanPRB16}.
Predictions from the SRI model show that a secondary shock wave will immediately merge with the leading shock wave at the upstream pore surface (see the location of the leading wave in Figure \ref{fig5} A).
In both the MD and SRD model, the leading shock wave has 'detached' from the upstream pore surface which means that the merger of the secondary shock wave from ejecta impact and leading wave will occur at some time delay after the pore has fully collapsed. 
In the limit of a strongly viscoplastic pore collapse event, the secondary shock will never catch up to the leading wave due to this time delay and the fact the ejecta impact generates a non-planar wave that decays in strength over its propagation distance\cite{bowden1952}. 

Characteristically different material flows around collapsing pores can be see as the strength of the shock is varied, and this was schematically represented as insets to Figure 1. 
From a set of MD simulations for 100nm pore diameters, we can confirm this behavior using specialized analysis tools available in the OVITO visualization code\cite{Stukowski2009}. 
The 'Atomic Strain' calculation within OVITO computes the local volumetric and von Mises shear strain for each atom based on the deformation tensor that maps the current atomic neighbor onto that of a given reference structure, which here we have chosen as the uncompressed initial condition of the simulation. 
Furthermore, only the carbon atoms were used to calculate these local strains because the hydrogens and nitro groups are subject to large amplitude vibrations that could be misinterpreted as strain with this method. 
A volume of material surrounding the pore equal to twice the area as the pore itself was selected for this analysis, it is subsequently broken into 32 equal angular segments. 
Average atomic strains within each of these angular segments is recorded for a snapshot just before the ejecta impacts the far pore wall (see Figure S1 A), it is the same relative time as in Figure 5). 
The zero angle in this plot is defined as the upstream edge (leftmost) of the pore and the strain is averaged over either hemisphere from this zero angle. 
Data in Figure S2 is displayed visually in panels B)-D) as the color of each segment, where areas of highest and lowest strain are displayed in red and blue. 
Strong shock waves, as seen in Figure S1 A), show a uniform strain field around the upstream surface with very little deformation at the downstream wall before the fluid-like ejecta arrives. 
Conversely, a more symmetric material flow is observed for the weakest shocks, as seen in Figure S1 D), that would could afford to run in MD.

In Section 5 it was discussed that the viscous heating around a collapsing pore contributes a significant amount of thermal energy to the resultant hot-spot, and this is particularly true for weak shocks resulting in viscoplastic pore collapse. 
To further demonstrate this, Figure S2 shows the thermal distribution and local temperatures for a 100nm pore shocked with a piston moving at 0.50km/s. 
Figure S2 B) displays the local temperatures at the last snapshot before ejecta impact; this temperature distribution is also captured in Figure S2 A) as the 'Pore Collapse Complete' series. 
At this shock pressure (~2GPa), the material that fills the void is moving very slow relative to the piston velocity and as such the ejecta impact is no longer a potent heat source for this hot-spot forming defect. 
This effect can be clearly seen as the limited area labeled 'Ejecta Impact' in Figure S2 A). 
In addition, there is little growth in the hot-spot after 15ps post collapse; the area designated 'Exothermic Reactions' has only grown to a few $nm^2$, and these exothermic reactions are critically needed to sustain the continued growth of the hot spot. 

\begin{figure}[]
\includegraphics{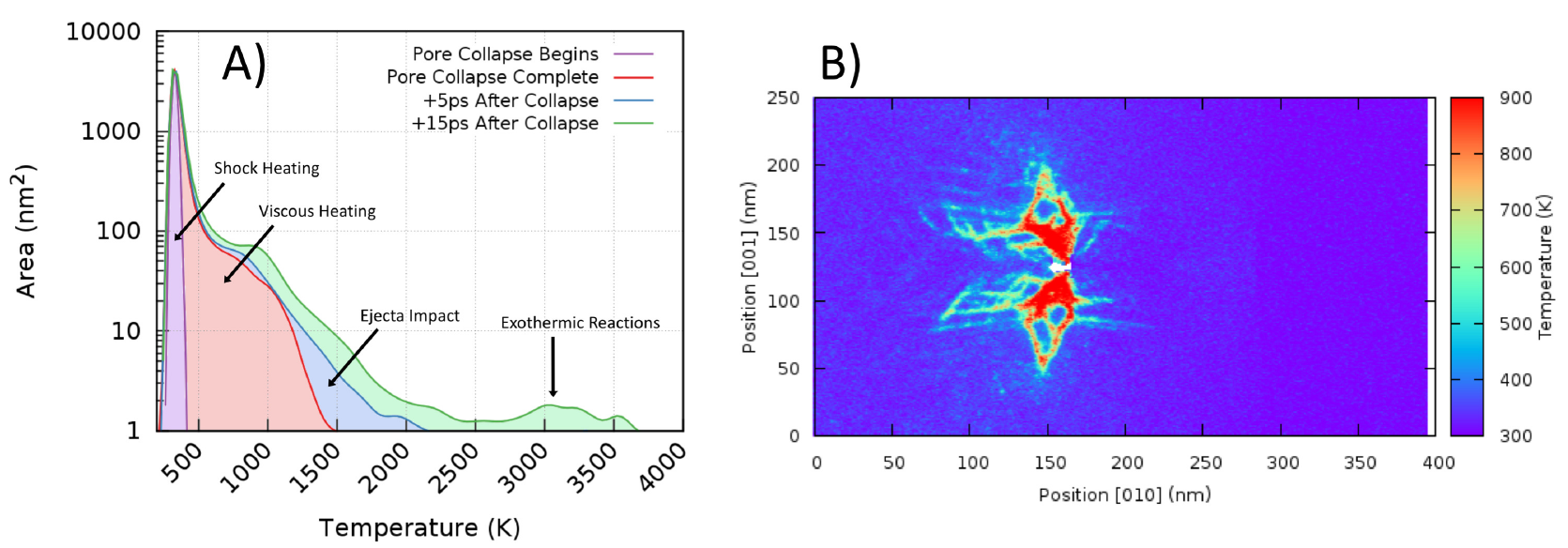}%
\caption{\label{Sfig2} {\bf A)} Temperature histograms for a few critical stages in the $U_{p}=0.50km/s$ shock response of HNS with a single 100nm void via ReaxFF MD. Arrow indicated regions exemplify the physical mechanism that predominantly contribute to that area of the temperature distribution. {\bf B)} Temperature field of the $U_{p}=0.50km/s$, 100nm pore simulation at last frame before ejecta impact, this corresponds to the \emph{Pore Collapse Complete} data in left panel.}
\end{figure}

\begin{thebibliography}{99}
\bibitem{sidecomment1}
If the time delay after ejecta impact is set to +15ps versus +5ps, the ratio of viscoplastic to jet impact heating is reduced to 7.5:1, 2:1, 1:1.33 and 1:1.27 for piston velocities of 0.5, 0.75, 1.0 and 1.25km/s, respectively. This, however, includes a much larger contribution from exothermic chemistry which is not entirely caused by jet impact.
\end{thebibliography}
\end{document}